\documentclass[twocolumn,showpacs,preprintnumbers,amsmath,amssymb]{revtex4-1}

\usepackage{graphicx}
\usepackage{dcolumn}
\usepackage{bm}
\usepackage{color}
\usepackage{ulem}

\begin{document}

\title{
Non-Kondo mechanism for resistivity minimum in spin ice conduction systems
}

\author{Masafumi Udagawa$^{1,2}$\thanks{E-mail address: udagawa@ap.t.u-tokyo.ac.jp}, Hiroaki Ishizuka$^1$, and Yukitoshi Motome$^1$}
\affiliation{%
$^1$Department of Applied Physics, University of Tokyo, Tokyo 113-8656, Japan\\
$^2$Max-Planck-Institut f\"{u}r Physik komplexer Systeme, 01187 Dresden, Germany}%

\date{\today}

\begin{abstract}
We present a mechanism of resistivity minimum in conduction electron systems coupled with localized moments, 
which is 
distinguished from the Kondo effect.
Instead of the spin-flip process in the Kondo effect, 
electrons are elastically scattered by local spin correlations which evolve 
in a particular way under geometrical frustration as decreasing temperature. 
This is demonstrated by the cellular dynamical mean-field theory for a spin-ice type 
Kondo lattice model on a pyrochlore lattice. 
Peculiar temperature dependences of the resistivity, specific heat, and magnetic susceptibility in the non-Kondo mechanism are compared 
with the experimental data in metallic Ir pyrochlore oxides. 
\end{abstract}

\maketitle
The interplay between conduction electrons and localized moments is one of the most fundamental subjects in 
strongly correlated electron systems. 
It leads to a variety of fascinating phenomena in electronic and magnetic properties. 
Recent intensive studies have revealed that anomalous magneto-transport properties emerge from the coupling to localized magnetic moments~\cite{Nagaosa,Tokura00}.

Kondo effect is a basic concept to understand the properties of electron-spin coupled systems.
As its simplest manifestation, the electrical resistivity shows a minimum as decreasing temperature ($T$) in a metal with magnetic impurities.
Kondo successfully explained the origin of resistivity minimum in terms of enhanced spin-flip scattering 
by magnetic impurities at low $T$~\cite{Kondo64}. 
After his pioneering work, the theory was elaborated to give a unified description of the system through a single energy scale, the Kondo temperature $T_{\rm K}$~\cite{Hewson93}.
 
\begin{figure}[t]
\begin{center}
\includegraphics[width=0.45\textwidth]{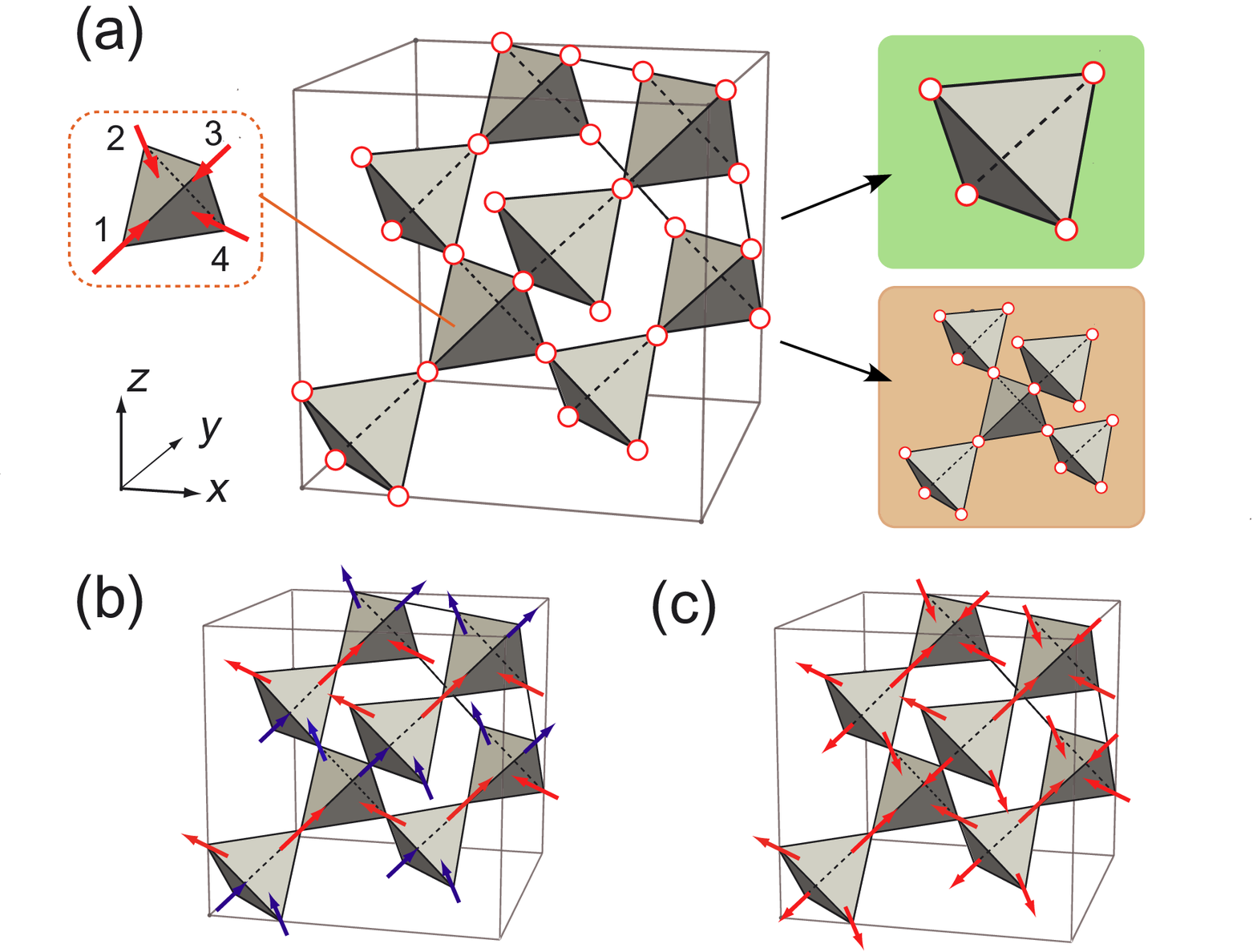}
\end{center}
\caption{\label{pyrochlore} 
(color online). (a) A cubic unit cell of a pyrochlore lattice. 
The left panel shows the four sublattices with the $\langle 111 \rangle$ easy axes. 
The right panels show 4- and 16-site clusters used in the CDMFT calculations.
(b) and (c) illustrate the magnetic ordering patterns for two-in two-out and all-in/all-out ordered states, respectively. 
}
\end{figure}

Recently, resistivity minimum has drawn a renewed interest since it was observed in several metallic Ir pyrochlore oxides, such as
Pr$_2$Ir$_2$O$_7$~\cite{Nakatsuji06} and Nd$_2$Ir$_2$O$_7$ under pressure~\cite{Sakata11}.
These compounds include conduction electrons in Ir $5d$ orbitals interacting with magnetic moments from rare-earth $4f$ localized electrons.  
The situation can be well characterized as a Kondo lattice system, and importance of quantum fluctuations in $4f$ moments was pointed out~\cite{Nakatsuji06,Machida07,Machida10,Balicas11,Onoda11}. 
Nevertheless, it is still unclear whether the Kondo scenario is straightforwardly applied to the observed resistivity minimum. 
Firstly, the local ground state of $4f$ multiplet is a non-Kramers doublet separated from excited states by a fairly large gap $\sim 160 $~K~\cite{Machida05}. This leads to a strong easy-axis anisotropy of $4f$ moments along the $\langle 111 \rangle$ directions in the pyrochlore lattice (see Fig.~\ref{pyrochlore}). 
The strong anisotropy considerably reduces $T_{\rm K}$, and hence, it is unclear whether the resistivity minimum at high $T \sim 40$~K is explained by the Kondo effect.
Secondly, the magnetic susceptibility shows a diverging behavior below $40$~K, in sharp contrast to a saturation in the Kondo effect~\cite{Nakatsuji06}. 
Thirdly, the specific heat shows a peak at $\sim 2$~K much lower than $40$~K~\cite{Nakatsuji06}, suggesting the absence of entropy release due to the Kondo effect around $40$~K.

In this Letter, we propose an alternative mechanism for the unconventional resistivity minimum. 
Instead of the Kondo effect, we focus on the role of anomalous spatial magnetic correlation imposed by geometrical frustration.
In these systems, it is inferred that localized moments develop spin-ice type local correlation~\cite{Nakatsuji06,Machida07}; that is, at four vertices in each tetrahedron in the pyrochlore lattice, 
two spins point inward and the other two point outward, without showing long-range ordering~\cite{Harris97,Ramirez99}. 
In the present study, we show that conduction electrons interacting with the spin-ice type moments exhibit a resistivity minimum at a characteristic temperature $T_{\rm min}$, below which the spin-ice correlation develops. 
The magnetic susceptibility and the specific heat also show characteristic $T$ dependences qualitatively different from those in Kondo impurity systems. 
The results are discussed in comparison with the experimental data for Pr$_2$Ir$_2$O$_7$ and Nd$_2$Ir$_2$O$_7$.

We start with a Kondo lattice model on a pyrochlore lattice (Fig.~\ref{pyrochlore}), whose Hamiltonian is given by
\begin{align}
\mathcal{H} = 
-t \sum\limits_{\langle i,j\rangle, s} (c_{is}^{\dag}c_{js} + {\rm H. c.}) 
- J\sum\limits_{i,s,s'} c_{is}^{\dag}{\bm\sigma}_{ss'}c_{is'}\cdot{\mathbf S}_i. 
 \label{eq:H}
\end{align}
Here, the first term describes the kinetic energy of electrons. The sum $\langle i,j\rangle$ is taken over the nearest-neighbor (n.n.) sites. 
The second term represents the coupling between conduction electron spins and localized moments ${\mathbf S}_i$ ($\bm{\sigma}$ is the Pauli matrix).
We consider the limit of strong easy-axis anisotropy along the $\langle 111 \rangle$ directions for the localized moments, namely, ${\mathbf S}_i$ is a classical Ising spin ($|{\mathbf S}_i|=1$) parallel to the local easy axis at each four-sublattice site, as shown in Fig.~\ref{pyrochlore}(a). 
In this limit, quantum fluctuation of the charge-spin coupling is completely suppressed:  the sign of $J$ does not matter, and the possibility of Kondo effect is excluded. 
Hereafter, we take $t=1$ as an energy unit, and the lattice constant of the 16-site cubic unit cell as a length unit. 

To study this model, we adopt the cellular dynamical mean-field theory (CDMFT)~\cite{Kotliar01}.
Through the previous studies of a double-exchange model, DMFT was shown to be a powerful tool for describing the physics of conduction electrons coupled to thermally fluctuating classical spins~\cite{Furukawa02}.
To take into account the peculiar spatial correlations under the geometrical frustration in the pyrochlore structure, we analyze the model (\ref{eq:H}) by using the CDMFT with clusters composed of $N_c = 4$ and $16$ sites as shown in Fig.~\ref{pyrochlore}(a).
We start with an initial Weiss field $g_{\eta\eta'}(i\epsilon_p)$, where $\epsilon_p = (2p+1)\pi T$ is the Matsubara frequency, and $\eta = (\alpha, s)$ takes
$2N_c$ values corresponding to the site index in each cluster $\alpha$ and spin $s$. The local Green's function is obtained by 
\begin{eqnarray}
\hat{G}(i\epsilon_p) = \langle\hat{G}^{\{{\mathbf S}_{\alpha}\}}(i\epsilon_p)\rangle\equiv{\rm Tr}_{\{{\mathbf S}_{\alpha}\}} \hat{G}^{\{{\mathbf S}_{\alpha}\}}(i\epsilon_p) P(\{{\mathbf S}_{\alpha}\}), 
 \label{impG}
 \end{eqnarray}
where $P(\{{\mathbf S}_{\alpha}\}) = e^{-S_{\rm eff}(\{{\mathbf S}_{\alpha}\})}/{\rm Tr}_{\{{\mathbf S}_{\alpha}\}}e^{-S_{\rm eff}(\{{\mathbf S}_{\alpha}\})}$ 
 with $e^{-S_{\rm eff}(\{{\mathbf S}_{\alpha}\})} = \prod_p{\rm det} [-\hat{G}^{\{{\mathbf S}_{\alpha}\}}(i\epsilon_p)]$, and $\beta=1/T$.
Here, $G_{\eta\eta'}^{{\{{\mathbf S}_{\alpha}\}}}(i\epsilon_p)$ is given by 
$\hat{G}^{{\{{\mathbf S}_{\alpha}\}}}(i\epsilon_p) = [\hat{g}^{-1}(i\epsilon_p) + J(\bm{\sigma}\cdot{\mathbf S}_{\alpha})\delta_{\alpha \alpha'}]^{-1}$ for each spin configuration $\{{\mathbf S}_{\alpha}\}$. (The hat indicates a $2N_c \times 2N_c$ matrix).
We calculate the trace in Eq.~(\ref{impG}) exactly by enumerating all $2^{N_c}$ spin configurations with the help of fast update algorithm~\cite{Gull}. 
Hence, we are free from any approximations or statistical errors in solving the impurity cluster problem.
A closed set of the self-consistent equations of CDMFT is given by Eq.~(\ref{impG}), the Dyson equation,  
$
\hat{\Sigma}(i\epsilon_p) = \hat{g}^{-1}(i\epsilon_p) - \hat{G}^{-1}(i\epsilon_p),
$
and the relation between the lattice and the impurity Green's function,
$
\hat{G}(i\epsilon_p) = \frac{1}{N}\sum_{\mathbf k}[{i\epsilon_p - \hat{t}({\mathbf k}) + \mu - \hat{\Sigma}(i\epsilon_p)}]^{-1}. 
$
Here $\hat{t}({\mathbf k})$ is the Fourier transform of the hopping matrix in Eq.~(\ref{eq:H}), $\mu$ is the chemical potential, and $N$ is the total number of clusters in the whole lattice. Dynamical quantities are calculated via
the retarded Green's function obtained simply by replacing $i\epsilon_p$ by $\epsilon + i\delta$ in Eq.~(\ref{impG}). 
The electrical conductivity is calculated by the Kubo formula with evaluating current-current correlation functions on the real-frequency axis.

\begin{figure}[t]
\begin{center}
\includegraphics[width=0.45\textwidth]{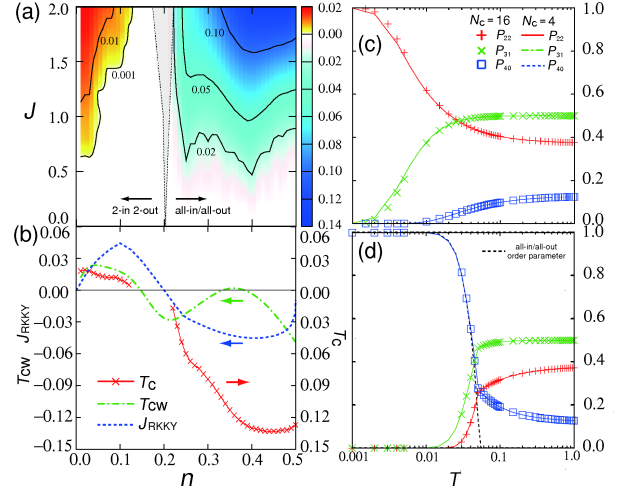}
\end{center}
\caption{\label{icity} 
(color online). (a) Particle density ($n$) and charge-spin coupling ($J$) dependences of the transition temperature ($T_c$). 
The thin dashed lines indicate the phase boundaries at $T=0$. The gray area shows the phase separated region.
(b) $n$ dependences of n.n. component of RKKY interaction ($J_{\rm RKKY}$), Curie-Weiss temperature ($T_{\rm CW}$), and $T_c$ at $J=2$. 
(c) and (d) show $T$ dependences of the probabilities for each spin configuration in a tetrahedron at $n=0.05$ and $0.40$, respectively: 
$P_{22}$, $P_{31}$, and $P_{40}$ correspond to the two-in two-out, three-in one-out or one-in three-out, and all-in or all-out configuration, respectively. 
The results of $N_c=4$ and $16$ are plotted at $J=1$.
The dashed curve in (d) shows the order parameter for the all-in/all-out state. }
\end{figure}

Figure~\ref{icity}(a) shows the phase diagram obtained by CDMFT. 
We focus on the density region less than half filling, $n = \sum_{i\sigma} \langle c_{i\sigma}^\dagger c_{i\sigma} \rangle / N N_c <0.5$. 
Hereafter, we assume a single tetrahedron as a magnetic unit cell~\cite{note_MC}. 
We obtain two dominant ground states in the present paramter range: One is a two-in two-out ordered state in the low-$n$ region [Fig.~\ref{pyrochlore}(b)], and the other is an all-in/all-out ordered state near half filling [Fig.~\ref{pyrochlore}(c)].
Comparison of the ground-state energies between two phases gives the phase boundary at $n\sim 0.2$ with a phase-separated region, as shown in Fig.~\ref{icity}(a).  
The gradation and contour lines represent the transition temperature $T_c$ for each ordering calculated by CDMFT with $N_c=4$~\cite{note}. 
The results are almost independent of the cluster size, as we will see below. 

\begin{figure}[t]
\begin{center}
\includegraphics[width=0.45\textwidth]{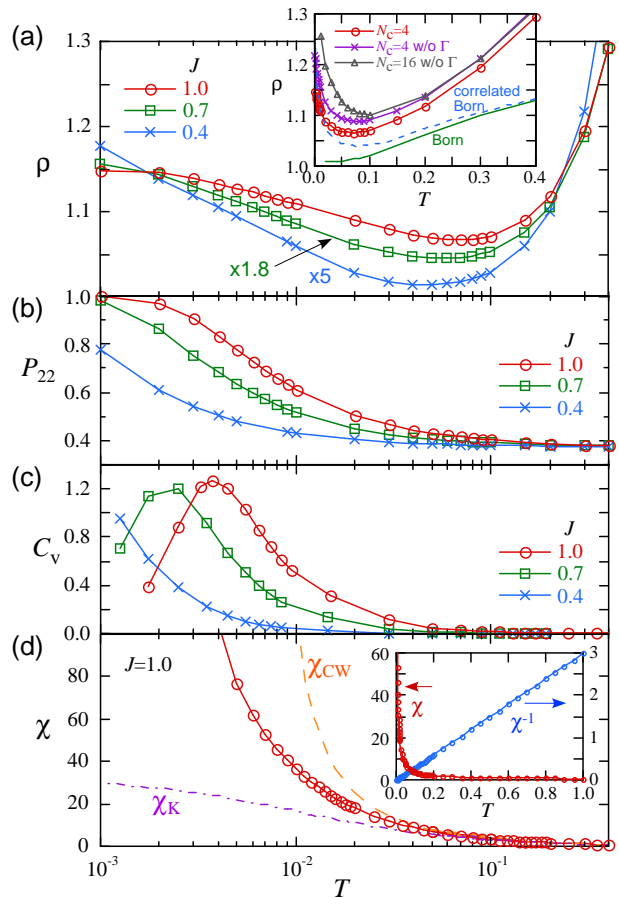}
\end{center}
\caption{\label{transport} 
(color online). $T$ dependences of (a) resistivity $\rho$, (b) $P_{22}$, (c) specific heat $C_v$, and (d) magnetic susceptibility $\chi$. 
The data are obtained at $n=0.05$ by CDMFT with $N_c=4$ cluster. 
$\rho$ is measured in unit of $h/8\pi e^2$, where $h$ is Planck constant and $e$ is elementary electronic charge. 
To facilitate the comparison, the data for $J=0.7$ and $0.4$ are multiplied by $1.8$ and $5.0$, respectively.
The inset of (a) shows the comparison of five different levels of approximation.
In (d), the Curie-Weiss fitting $\chi_{\rm CW}$ and ``Kondo" susceptibility $\chi_{\rm K}$ are shown for comparison.
The inset shows the plot of $\chi$ and $\chi^{-1}$ in the linear $T$ scale. 
}
\end{figure}

In order to understand the phase diagram, we compare the n.n. component of the Ruderman-Kittel-Kasuya-Yosida (RKKY) interaction, $J_{\rm RKKY}$~\cite{RKKY}, the Curie-Weiss temperature $T_{\rm CW}$, and $T_c$ in Fig.~\ref{icity}(b) for $J=2$. 
Here, $J_{\rm RKKY}$ is calculated from the eigenstates of Hamiltonian (\ref{eq:H}) at $J=0$, by the second-order perturbation in terms of $J/t$. 
$T_{\rm CW}$ is obtained by the Curie-Weiss fit for the magnetic susceptibility as $\chi^{-1} \propto (T-T_{\rm CW})$ in the range of $0.5<T<1.0$ [see also Fig.~\ref{transport}(d)]~\cite{calc_chi}. 
The ground state is well understood from $J_{\rm RKKY}$. $J_{\rm RKKY}$ is ferromagnetic in the low-$n$ region, reflecting a small Fermi surface. 
The ferromagnetic interaction between neighboring localized moments favors the two-in two-out configuration. As $n$ is increased, however, $J_{\rm RKKY}$ changes its sign at $n\sim 0.2$ and becomes antiferromagnetic, which favors the all-in/all-out configuration appearing for $n > 0.2$. 
On the other hand, $T_{\rm CW}$ has less correlation with the ordering pattern. 
This may be attributed to the long-range and oscillating nature of RKKY interaction~\cite{IkedaKawamura}. 

Although the sign of $J_{\rm RKKY}$ well accounts for the dominant phases at $T=0$, the magnitude of $J_{\rm RKKY}$ does not explain large difference of $T_c$ between two phases. 
$T_c$ for the two-in two-out ordering is largely suppressed compared to that for the all-in/all-out ordering, despite the magnitude of $J_{\rm RKKY}$ is similar in the two regions, as shown in Fig.~\ref{icity}(b). 
The contrasting $T_c$ may be a manifestation of the frustration effect.  
Namely, ferromagnetic $J_{\rm RKKY}$ is underconstraint and results in macroscopic degeneracy among different configurations of two-in two-out tetrahedra, as is well known in spin ice~\cite{Harris97,Ramirez99}: A true long-range ordering is driven by further-neighbor RKKY interactions, which are smaller than the n.n. $J_{\rm RKKY}$.  
On the other hand, the all-in/all-out configuration is free from the frustration and a unique long-range order is selected by the n.n. antiferromagnetic $J_{\rm RKKY}$ alone. 
Thus, the geometrical frustration plays an important role on the suppression of two-in two-out ordering in the low-$n$ region. 

The suppression of $T_c$ leads to emergence of peculiar liquid-like behavior above $T_c$. 
In Figs.~\ref{icity}(c) and \ref{icity}(d), we plot the weights of local spin configurations within a tetrahedron for $n=0.05$ and $0.40$ at $J=1$, respectively. $P_{22}$, $P_{31}$, and $P_{40}$ are the probabilities that we find two-in two-out, three-in one-out or one-in three-out, and all-in or all-out configurations in a tetrahedral unit, respectively. The results of $N_c=4$ and $16$ coincide with each other to a remarkable extent. 
For $n=0.40$, a transition with all-in/all-out ordering occurs at $T_c\simeq 0.05$, and accordingly, $P_{40}$ shows a kink-like singularity and increases rapidly below $T_c$. 
On the other hand, for $n=0.05$, $T_c$ is suppressed below $0.001$, despite that $P_{22}$ starts to increase considerably at $T \sim 0.1$ and exceeds 0.999 at $T=0.001$. 
Thus, the two-in two-out spin-ice type local correlation is developed from much higher $T$ than $T_c$, in sharp contrast to the all-in/all-out case. 
The growth of $P_{22}$ without long-range ordering is a clear signature of emergence of spin liquid state. 

We find that the peculiar liquid state exhibits an interesting transport property. 
Figure~\ref{transport}(a) shows $T$ dependence of the electrical resistivity $\rho$ at $n=0.05$ for three different values of $J$, obtained with $N_c=4$. 
The resistivity exhibits a clear minimum at $T\equiv T_{\rm min}$ and subsequent upturn below $T_{\rm min}$. 
The resistivity upturn is somewhat counterintuitive. 
Usually, in electron-spin coupled systems, magnetic correlations develop at low $T$ so as to lower the kinetic energy and restore the coherent motion of electrons, which leads to the reduction of resistivity. 
The mechanism is widely seen in a ferromagnetic ordering of the double-exchange systems~\cite{Furukawa02}. 
Then, the question is what brings about the resistivity upturn below $T_{\rm min}$ in the present pyrochlore system.

In order to gain an insight into the mechanism, we compare $\rho$ calculated at several different levels of
approximation. 
The results are shown in the inset of Fig.~\ref{transport}(a).  
In the first place, $\rho$ shows only a small change by ignoring the vertex correction $\Gamma$, which
implies that the self-energy effect or enhanced scattering rate plays a central role in the resistivity minimum. 
To clarify the effects of spin correlations via the self-energy, we use the perturbation up to the second order in terms of $J/t$ as
$\Sigma_{\eta\eta'} = J^2 g_{\eta\eta'}\langle{\mathbf S}_{\alpha}\cdot{\mathbf S}_{\alpha'}\rangle$.
When neglecting spatial spin correlation in $\hat{\Sigma}$ by taking $\langle{\mathbf S}_{\alpha}\cdot{\mathbf S}_{\alpha'}\rangle = \delta_{\alpha\alpha'}$, $\rho$ decreases monotonically as lowering $T$. 
(This corresponds to the second-order Born approximation.)
Alternatively, when we use the CDMFT result of $\langle{\mathbf S}_{\alpha}\cdot{\mathbf S}_{\alpha'}\rangle$ in $\hat{\Sigma}$ to incorporate the effect of growing spin ice correlation (we call this scheme the ``correlated Born approximation"), $\rho$ shows clear upturn, qualitatively consistent with the nonperturbative CDMFT result. 
This striking difference by considering short-range spin correlations corroborates the crucial importance of spin ice correlation in the resistivity upturn. 
Indeed, $T_{\rm min}$ roughly corresponds to the onset of two-in two-out correlation $P_{22}$, as shown in Figs.~\ref{transport}(a) and \ref{transport}(b).
We also note that the result for $N_c=16$ gives a steeper upturn than that for $N_c=4$, 
suggesting that spatial correlations beyond the n.n. sites enhance the scattering of electrons.

The above considerations clearly show that 
the resistivity minimum in the present system is caused by the peculiar local spin correlation growing under the strong geometrical frustration.
Here, the lowering of conduction electron energy gives rise to the enhancement of two-in two-out local spin correlations, leading to a formation of spatially disordered configurations of two-in two-out tetrahedra. 
Our results indicate that the ``spin-ice manifold" acts as a rather stronger scatterer for electrons, than simple paramagnetic moments. 
Below $T_c$, two-in two-out tetrahedra are aligned and the degeneracy of the manifold is lifted, leading to a sharp drop of $\rho$. 
Similarly, an application of magnetic field above $T_c$ also causes a reduction of $\rho$, i.e., negative magnetoresistance (not shown).  
The coexistence of seemingly contradicting features --- reduction of conduction electron energy and enhanced scattering rate --- is indeed a manifestation of 
unusual physics in the itinerant frustrated system. 

Now we discuss $T$ dependence of thermodynamic quantities. 
Figure~\ref{transport}(c) shows the specific heat $C_v\equiv T \partial {\cal S} / \partial T$, calculated from
the spin entropy ${\cal S}\equiv -\sum_{\{{\mathbf S}_{\alpha}\}}P(\{{\mathbf S}_{\alpha}\})\log P(\{{\mathbf S}_{\alpha}\})$
\cite{ElectronicContribution}.
The results are obtained with $N_c=16$. 
$C_v$ does not have any significant feature at $T \sim T_{\rm min}$, but shows a broad peak at a much lower $T$. 
The peak temperature roughly corresponds to the inflection point of $P_{22}$ in Fig.~\ref{transport}(b), suggesting that the entropy is released at the saturation of spin ice correlation, not at the onset. 
On the other hand, the magnetic susceptibility $\chi$ continuously increases as $T$ decreases, as shown in Fig.~\ref{transport}(d). 
The high-$T$ behavior is well fitted by the Curie-Weiss law, 
$
\chi_{\rm CW}\equiv C /(T - T_{\rm CW}) 
$ [see inset of Fig.~\ref{transport}(d)]. 
$T_{\rm CW}$ depends on $n$ as well as $J$, as shown in Fig.~\ref{icity}(b). 
As lowering $T$, $\chi$ deviates from $\chi_{\rm CW}$ below $T_{\rm min}$, while it continues to diverge weaker than $\chi_{\rm CW}$, as shown in Fig.~\ref{transport}(d). 
For comparison, we plot a mimic of the high-$T$ susceptibility of the Kondo system, $\chi_{\rm K} = C/(T+T_{\rm K})$, with taking a small enough $T_{\rm K} = 0.01$.
The diverging $\chi$ as well as the broad peak of $C_v$ at low $T$ is in marked contrast with the canonical Kondo behaviors. 

Finally, let us discuss our results in comparison with the experimental data of Pr$_2$Ir$_2$O$_7$~\cite{Nakatsuji06} and Nd$_2$Ir$_2$O$_7$ under pressure~\cite{Sakata11}.
Our scenario of resistivity minimum due to the growing spin-ice local correlation explains many experimental features, at least qualitatively, such as the large separation of $T$ scales between the resistivity minimum and specific heat peak, and the diverging behavior of the magnetic susceptibility~\cite{Nakatsuji06}. 
These features are expected to remain robust down to a smaller realistic value of $J/t$. 
They will also be unaltered even when the exchange interaction between localized moments is included in the model, 
which is small but a finite in real compounds. 
Compared to the canonical Kondo effect, our scenario is free from the difficulties mentioned in the introduction, and therefore, potentially gives a unified explanation for the resistivity upturn in experiments. 
It also naturally explains the resistivity drop by long-range ordering at a lower $T$ than $T_{\rm min}$ observed in Nd$_2$Ir$_2$O$_7$ under pressure~\cite{Sakata11}.
A smoking-gun experiment is a photoemission spectroscopy; a Kondo resonance peak, which is naturally expected to develop below $T_{\rm min}$ in the canonical Kondo systems, is absent in our scenario. 

In summary, we have studied the transport properties of conduction electrons interacting with spin-ice type moments. We found that the resistivity takes a minimum at a characteristic temperature, below which local spin-ice correlations begin to grow under geometrical frustration. 
Our results agree with the experimental results of Pr$_2$Ir$_2$O$_7$ and Nd$_2$Ir$_2$O$_7$ in many respects,
supporting our scenario as a promising mechanism of resistivity upturn in these compounds. 
The family of Ir pyrochlores exhibits metal-insulator transitions 
by chemical substitutions~\cite{Matsuhira07,Matsuhira11} or external pressure~\cite{Sakata11,Tafti11}, 
in which Coulomb repulsion and spin-orbit interaction may play an important role. 
Furthermore, in Pr$_2$Ir$_2$O$_7$, an anomalous Hall effect without magnetic ordering was reported at very low $T$, and the importance of quantum fluctuations of Pr moments was discussed~\cite{Machida07,Machida10,Balicas11,Onoda11}. 
It is interesting to extend the present study for comprehensive understanding of these phenomena
by including these elements neglected in our model.
It is also intriguing to ask how our scenario is universal. 
Other types of spatial fluctuations such as orbital and lattice degrees of freedom, may lead to resistivity upturn as well as other fascinating transport phenomena under severe frustration. 
The present results give a firm ground for such extensions.  

The authors thank K.\ Matsuhira, S.\ Nakatsuji, R.\ Moessner, and P.\ A.\ McClarty for fruitful discussions.
H.I. is supported by Grant-in-Aid for JSPS Fellows.
This work was supported by KAKENHI (Nos. 19052008, 21340090, 21740242, 21340090, and 22540372), 
Global COE Program ``the Physical Sciences Frontier," 
and the HPCI Strategic Program, from MEXT, Japan.


\begin{thebibliography}{9}
\bibitem{note_MC} We also performed a real-space Monte Carlo simulation. The result confirms that the two ordered states in Fig.~\ref{icity}(a) appear, whereas some other phases with more complicated magnetic orderings intervene in the intermediate $n$ region. The details will be reported elsewhere.
\bibitem{note} Our model should exhibit some magnetic ordering in the low-$T$ limit, and hence, the white region in the phase diagram simply indicates that $T_c$ is smaller than $0.001$.
\bibitem{calc_chi} $\chi$ is obtained from the magnetization by applying a small magnetic field $h=0.001$ in the $z$ 
direction only for Ising moments. No dependence on the field direction is observed for this small $h$.
\bibitem{ElectronicContribution} Electronic contribution to the specific heat is negligibly small in this $T$ range, e.g., $\sim 0.01$ at $T=0.01$ for $J=1.0$.
\end{thebibliography}
\end{document}